# Optimal Probing Sequences for Polarization-Multiplexed Coherent Phase OTDR

**Christian Dorize[1], Elie Awwad[1,2], Sterenn Guerrier[1] & Jérémie Renaudier[1]**
*(1) Nokia Bell Labs France, Route de Villejust, 91620 Nozay, France; (2) Télécom Paris, 19 place Marguerite Perey, 91120 Palaeseau, France*
christian.dorize@nokia-bell-labs.com

**Abstract:** We introduce dual-polarization probing codes based on two circularly shifted frequency sweep signals enabling perfect channel estimation. This is achieved with a probing length equal to at least twice the fiber round-trip propagation time.

## 1. Introduction

The emergence of distributed fiber sensing exploiting the optical phase variations [1] has opened the way to a wide range of new applications where fibers capture low energy mechanical perturbations over a wide bandwidth. Moreover, the recent introduction of a dual polarization probing technique along with a coherent receiver where the fiber is jointly probed on two orthogonal polarization axes by means of mutually orthogonal codes derived from Golay sequences has shown to enable perfect estimation of the retro-propagated optical field [2]; polarization fading issues are solved, leading to substantial sensitivity gain for practical applications. The technique is suitable for sensing dedicated fibers as those which include equally spaced Fiber Bragg Gratings (FBGs) to reinforce the backscattered optical intensity. However, when probing standard Single Mode Fibers (SMF) deployed in metropolitan optical networks and submarine long-haul transmission links, the back-propagated field originates from the sole Rayleigh backscatter with reflectors randomly distributed along the fiber. A perfect channel estimation was shown to be also achievable with binary codes in this latter case, subject to the use of code lengths greater than 4 times the round-trip in the fiber to sense. This constraint reduces the maximal amount of probing codes transmitted per time unit and so the maximal achievable mechanical bandwidth by a factor of four compared with the theory.

This paper introduces alternative probing codes based on phase generated frequency sweeps leading to perfect channel estimation with a maximal bandwidth of half the theoretical limit (instead of one fourth with binary codes). Section 2 details the code design and compares it to the former binary approach based on Golay sequences whereas Section 3 compares the performance between the two approaches on a simulated SMF optical line thanks to a dual-polarization model of the Rayleigh backscatter optical field.

## 2. Design of dual-polarization codes

The probing signal choice in optical sensing has a strong impact on the performance and this is even more crucial in phase OTDR systems. Basic light pulse interrogation potentially induces optical non-linearities whereas standard sweep excitation on a single polarization axis is subject to polarization fading effects. To circumvent these effects, we introduced in [2] a probing technique based on mutually orthogonal binary codes derived from Golay sequences that simultaneously modulate two orthogonal polarization states of a continuous light signal. A dual-polarization coherent receiver, as usually used in long haul communication systems, captures the backpropagated optical field. This polarization-diversity interrogation technique, that can be named as Multiple Input Multiple Output (MIMO) since the polarization diversity applies to both the transmitter and the receiver sides, was shown to provide a perfect estimation of the probed fiber backscattered optical field. We recall below the theory when probing an SMF modeled with randomly distributed scattering points in its core.

After correlation of the received signals with the original ones, the backscattered signal vectors $\boldsymbol{E_r} = \begin{pmatrix} E_{rx} \\ E_{ry} \end{pmatrix}$ from the $i$<sup>th</sup> segment are given by $\boldsymbol{E_{r_{i,j}}} = \boldsymbol{H_{i,j}} \boldsymbol{E_t}$ where $\boldsymbol{E_t} = \begin{pmatrix} E_{tx} \\ E_{ty} \end{pmatrix}$ are the transmitted symbol-vectors onto each of two orthogonal polarization axes $x$ and $y$, and $\boldsymbol{H_{i,j}}$ is the dual-pass impulse response, represented by a 2×2 Jones matrix, up to the $i$<sup>th</sup> fiber segment at time index $j$. The absolute optical phase is extracted from each matrix $\boldsymbol{H_{i,j}}$ as $\boldsymbol{\varphi_{i,j}} = 0.5\angle \det(\boldsymbol{H_{i,j}})$ where det(.) stands for the determinant of $\boldsymbol{H_{i,j}}$. The computed absolute phases hold the phase evolution from the interrogator up to the $i$<sup>th</sup> segment; we obtain the phase evolution per segment by extracting the differential phases with the phase from the first reflector set as a reference.

The estimation problem is to find $E_t$ that allows for a perfect estimate of all matrices $H_{i,j}$ from the captured optical field. The initial solution proposed in [2] consists in probing the optical line with two mutually orthogonal complementary Golay binary pairs {$G_{a1}$, $G_{b1}$} and {$G_{a2}$, $G_{b2}$} that jointly modulate the two orthogonal polarization states through a Polarization-Division-Multiplexed (PDM) binary phase-shift-keying (BPSK, or 2-PSK) mapping. The complementary pairs are designed [3] by recursion from a 4-symbol seed of alphabet {-1,+1} to get the desirable length, yielding a probing code duration $T_{Golay} = 2.(4.2^K)/F_{symb}$, where $K$ is an integer standing for the number of recursions and $F_{symb}$ is the fixed symbol rate. When continuously probing the fiber with the period $T_{Golay}$, the line is analyzed within a mechanical bandwidth $BW=1/(2T_{Golay})$, whereas the spatial resolution $S_r=c_f/(2F_{symb})$ is adjusted through the symbol rate, $c_f$ standing for the light velocity in the fiber core.

The processing at the reception side consists of four correlations between each of the two received optical fields {$\boldsymbol{E_{rx}}, \boldsymbol{E_{ry}}$} and each of the two transmitted codes {$\boldsymbol{E_{tx}}, \boldsymbol{E_{ty}}$}, leading to periodical estimates of $H_{i,j}$. The condition for perfect estimation of $H_{i,j}$ from $E_r$ when the sensed fiber is probed with PDM-BPSK codes was shown [2,4] to be $T_{Golay}>4T_{ir}$, where $T_{ir}=2L/c_f$ stands for the time spreading of the channel response for a fiber of length $L$. Notice that selecting $T_{Golay}$ close to the lower limit $4T_{ir}$ maximizes the mechanical bandwidth $BW$ of the sensing system. The PDM-QPSK alternative mapping developed in [2] does not guarantee a noise-free estimation for the random Rayleigh backscattering intensity profile but is better suited for a periodical intensity profile as the one obtained from equally-spaced high-reflecting FBG arrays.

As an alternative to the above-mentioned probing technique, we consider phase-modulated CAZAC (Constant Amplitude Zero-Autocorrelation Code) sequences usually used in the telecommunication domain to perform channel estimation of a transmission link. The symbols from a class of perfect-squares minimum-phase CAZAC sequences [5] are defined as:

$$c_n = \exp\left(j\frac{2\pi}{\sqrt{N}}.\left(mod(n-1,\sqrt{N})+1\right).\left(\left\lfloor\frac{n-1}{\sqrt{N}}\right\rfloor+1\right)\right) \quad (1)$$

$n = \{1, ..., N\}$ denotes the time index of the symbol with $N = 4^M$, $M$ being a non-zero positive integer. The alphabet is composed of $2^M$ complex symbols having the same module and uniformly distributed over the unit circle, leading to a $2^M$-symbol Phase Shift Keying ($2^M$-PSK) constellation. Exploiting the properties of CAZAC sequences, we apply a circular shift of N/2 symbols onto the initial N-symbol CAZAC sequence **c** which leads to a sequence **c'**. Sending **c** and **c'** simultaneously over two polarization states offers mutual orthogonality [5] and achieves perfect channel estimation as for the above codes derived from Golay, however with a less-stringent constraint on the length of the probing sequence as demonstrated in Fig.1 below. When estimating a fiber Rayleigh backscattering over a round-trip duration of $T_{ir}$, the perfect estimation is now obtained under the condition $T_{cazac}>2T_{ir}$. Fig.1 shows a comparison of the backscattered optical field intensity estimated when the line is probed with the initial Golay codes in Fig.1(a) and with CAZAC ones in Fig.1(b) of the same length (M=7). A dual-polarization Rayleigh backscatter model [6] is used to simulate the backpropagated field from an 8.5km length SMF with an attenuation of 0.2dB/km. The line is interrogated with the shortest compatible PDM-BPSK Golay code, that is $T_{Golay}=4T_{ir}=0.3$ms. Fig.1(a) displays one period $T_{Golay}$ of the estimated backscattered intensity response. The actual fiber response spreads over the first displayed quarter on the left and is immediately followed by a spatial aliasing pattern. The response captured when interrogating the simulated optical fiber with the CAZAC sequences of the same length $T_{cazac}=T_{Golay}=4T_{ir}$, appears in Fig.1(b). As for Fig.1(a), the useful response part appears in the first quarter but the spatial aliasing pattern is now a copy of this response, and is concentrated in the center of the $4.T_{ir}$-long displayed window.

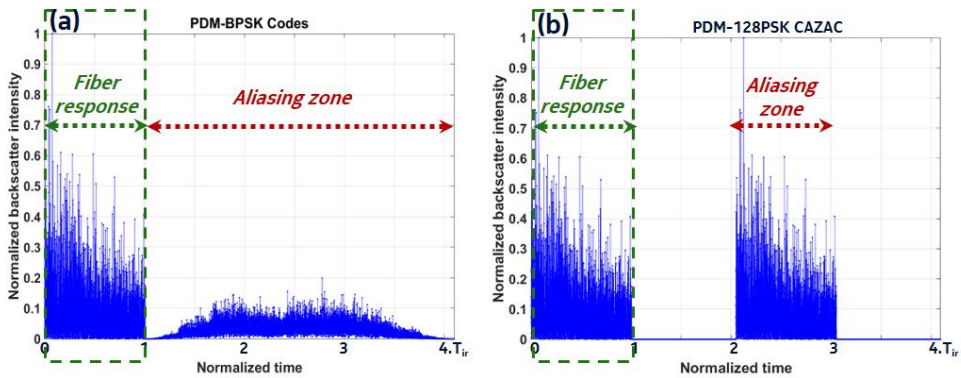

*Fig.1: Intensity response highlighting PDM-BPSK(a) and PDM-CAZAC(b) codes respective aliasing patterns*

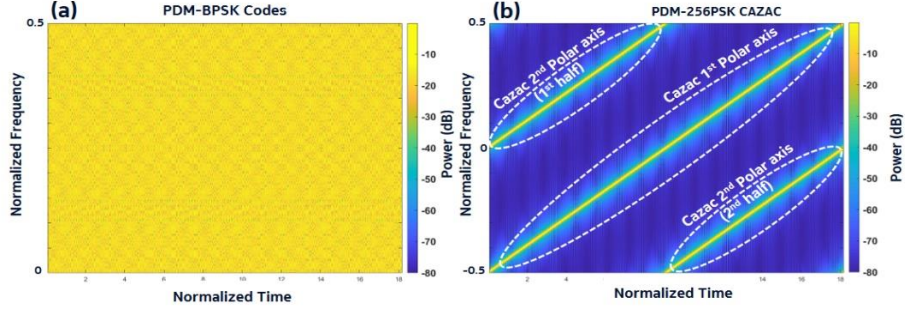

*Fig. 2: Energy distribution in the time-frequency planes of PDM-BPSK(a) versus PDM-CAZAC(b) codes*

The blank zone that separates the useful response from the spatial aliasing pattern highlights that we have not reached the maximal fiber probing distance with CAZAC codes; the distance can be extended up to a factor of 2 to reach the error-free estimation condition $T_{cazac}>2.T_{ir}$, whereas selecting $T_{Golay}$ smaller than $4.T_{ir}$ would lead to a fiber response estimation corrupted by aliasing.

A key difference between these two MIMO interrogation methods is illustrated through the respective projection of the probing sequences in a time versus frequency plane. Fig.2(a) and (b) display, for each of the two compared techniques their time-frequency signatures (one per polarization axis). They are obtained from a series of short-term spectral decompositions from sequences generated to have a length of $N=2^{16}$ symbols for both methods (M=8 with CAZAC). For PDM-BPSK codes, we observe a uniformly distributed energy and it is not possible to distinguish between the two sequences used to probe each polarization axis. Conversely, for the PDM-256PSK CAZAC case, the time-frequency plane highlights that the 2 CAZAC codes used to probe the optical channel onto both polarization axes are nothing but two linear frequency sweeps delayed by half the code period. This observation clarifies how the orthogonality property is achieved here: the two polarization axes are probed at each time instant with two distinct instantaneous frequencies, separated by $f_{Symb}/4$.

The frequency sweep interrogation, known as OFDR, is a well spread method for channel estimation in various application fields. Therefore, it is worth considering an alternative, somehow more intuitive, way to generate a dual-polarization frequency sweep as that achieved with the CAZAC method. A first amplitude modulated real sweep signal is used to linearly and uniformly cover the $[0:F_{symb}/2]$ bandwidth over a period $T_{sweep}=T_{cazac}$ over the first polarization tributary and a second identical signal, but delayed by half the code period, modulates simultaneously the second tributary. The associated signature in a time-frequency plane is equivalent to that shown in Fig.2(b). The two continuously repeated signals are, for a practical use case example, digitally generated and then converted by two Digital to Analog Converters (DACs) prior to modulating the laser light onto the two polarization axes. The next section examines the performance of the three above defined methods when probing simulated fibres of various length.

## 3. Simulation results

The backpropagated optical field from an SMF of length $L$ is simulated using our dual-polarization Rayleigh backscatter model [6]. The symbol rate $F_{symb}$ and the probing sequence length are fixed to *50M*baud and $2^{14}$ symbols respectively, yielding a sequence duration *T=328µs*, identical for the three interrogation methods. The associated spatial resolution is $S_r=2m$ ($c_f \approx 2.10^8 m/s$ in the SMF core) and the Rayleigh backscatter model generates one Jones matrix of the backpropagated field for each $S_r$ long segment along the fibre. The set of generated Jones matrices $H_i$ ($1<i<L/S_r$) is used as reference to compare with the estimated matrices yielded by each probing technique. We consider two error criteria per segment: the relative error on the matrix determinant and the absolute error on the phase. The maximal fiber length to be probed with PDM-BPSK codes to achieve perfect estimation is $L_{Code}^{Max} = \frac{1}{4} \frac{N_{Code} \cdot c_f}{2 \cdot f_{Symb}}$, which yields 8.2km here, whereas the maximal length $L_{Cazac}^{Max} = \frac{1}{2} \frac{N_{Cazac} \cdot c_f}{2 \cdot f_{Symb}}$ is twice larger when using CAZAC (or sweep) sequences with $N_{cazac}=N_{Codes}$. Fig.3(a) shows the mean relative error of the estimated Jones matrix determinants and Fig.3(b) displays the mean phase error as a function of the fiber length for each of the three considered probing methods. We generated the reference Jones matrices for a 20km long SMF and then we simulated the sensing over increasing portions of it, successively using the three probing techniques. The mean error is obtained by averaging the error per fiber segment up to the distance portion of interest. No laser phase noise is considered here, only an additive white Gaussian noise is added at the receiver to simulate thermal noise. Results in Fig.3(a) and (b) highlight the perfect channel estimation provided by the initial code interrogation technique up to the 8.2km limit. As expected, the error starts increasing beyond 8.2km whereas CAZAC codes keep providing a low estimation error until 16.4km.

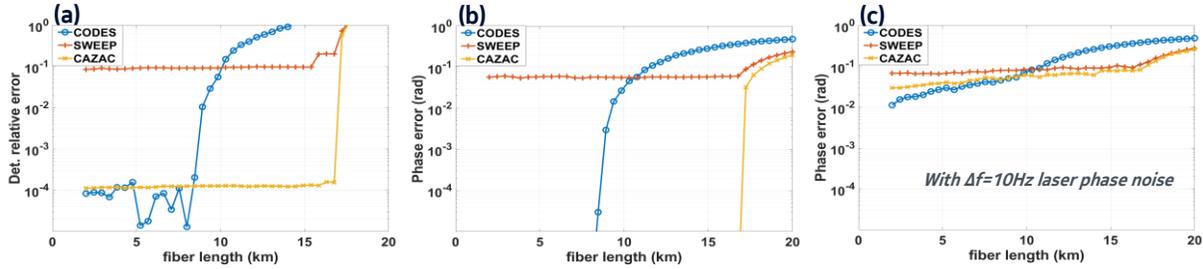

*Fig. 3: relative error on the Jones matrix determinant(a), phase error without(b) & with(c) laser phase noise*

Beyond this limit, CAZAC estimation error becomes immediately large, which is a consequence of its larger aliasing pattern with an energy concentrated over a much shorter zone when compared to the aliasing obtained for BPSK Golay codes, as shown in Fig.1. The continuously modulated (amplitude) sweep signal behavior shows an error floor prior to reaching the 16.4km limit. Though orthogonality between the two amplitude and frequency modulated sweep signals is fulfilled in theory, we conjecture the residual error is induced by the hard transition when switching from maximal instantaneous frequency $F_{symb}/2$ down to 0 within one symbol period. Conversely, the 2 frequency sweep signals from the CAZAC sequences are derived from phase shifts through a constant amplitude modulation, allowing to handle in a softer way the transitions between consecutive codes, with a lesser impact when crossing the instantaneous intermediate frequency $F_{symb}/4$ that modulates the orthogonal polarization axis.

We complement the study by simulating the imperfection of a laser source since the coherence loss is known to have a major impact on the phase estimation in coherent phase OTDR systems [6]. Fig.3(c) shows the estimated phase error with a laser linewidth *df=10Hz* simulated from a Lorentzian model. A comparison with the laser noise free performance in Fig.3(b) shows a significant growth of the phase estimation noise floor for both codes and CAZAC probing cases. The intrinsic noise floor induced by the dual-polarization sweep case remains beyond this limit.

Therefore, perfect channel estimation is verified for both binary PDM-BPSK and PDM-CAZAC cases, under their respective round-trip probing distance limit, which is twice longer with CAZAC sequences. When probing a fiber having the maximal allowed length with PDM-BPSK codes, that is *L*=8.2km with the above defined set of parameters, the maximal achievable bandwidth is $BW_{Code}^{Max} = \frac{1}{4}\frac{c_f}{4.L} = 3$kHz with BPSK whereas CAZAC sequences allow for $BW_{Cazac}^{Max} = \frac{1}{2}\frac{c_f}{4.L} = 6$kHz.

It can be noticed from Fig.3c that PDM-BPSK codes look more robust to laser phase noise below this 8.2km limit. This might be induced by the better frequency diversity over time provided by these codes, see Fig.2. Let's also point out that, from a practical point of view, PDM-CAZAC sequences require a more complex and accurately tuned transmitter to generate a $2^M$-PSK constellation for large values of M.

### 4. Conclusion

We demonstrated that a class of CAZAC sequences is an alternative candidate to previously introduced PDM-BPSK Golay codes for jointly probing two orthogonal polarization axes in Phase-OTDR. Both methods yield perfect estimation of the Rayleigh backscattered response, but the CAZAC-based method achieves it with a mechanical bandwidth twice larger than that of the initial codes. A PDM version of the more usual sweep signal probing technique (OFDR) was also studied but it was shown to exhibit a severe estimation noise. The existence of alternative sequences enabling perfect channel estimation of the round-trip Jones matrices after Rayleigh backscattering with a higher mechanical bandwidth than the proposed CAZAC solution remains an open question.